\documentclass[twocolumn,english,showpacs,floatfix,amsmath,amssymb,prl,superscriptaddress]{revtex4}
\usepackage[T1]{fontenc}
\usepackage[latin9]{inputenc}
\usepackage{babel}

\usepackage{graphicx}
\usepackage{amssymb}
\usepackage{esint}
\usepackage{hyperref}



\begin{document}

\title{Energy Partitioning of Tunneling Currents into Luttinger Liquids}

\author{Torsten Karzig} 
\affiliation{\mbox{Dahlem Center for Complex Quantum Systems and Fachbereich Physik, Freie Universit\"at Berlin, 14195 Berlin, Germany}}

\author{Gil Refael}
\affiliation{Department of Physics, California Institute of Technology, Pasadena, California 91125, USA}

\author{Leonid I. Glazman} 
\affiliation{Department of Physics, Yale University, 217 Prospect Street,
New Haven, Connecticut 06520, USA}

\author{Felix von Oppen}
\affiliation{\mbox{Dahlem Center for Complex Quantum Systems and Fachbereich Physik, Freie Universit\"at Berlin, 14195 Berlin, Germany}}

\date{\today}
\begin{abstract}
Tunneling of electrons of definite chirality into a quantum wire creates counterpropagating excitations, carrying both charge and energy. We find that the partitioning of energy is qualitatively different from that of charge. The partition ratio of energy depends on the excess energy of the tunneling electrons (controlled by the applied bias) and on the interaction strength within the wire (characterized by the Luttinger liquid parameter $\kappa$), while the partitioning of charge is fully determined by $\kappa$. Moreover, unlike for charge currents, the partitioning of energy current should manifest itself in $dc$ experiments on wires contacted by conventional (Fermi-liquid) leads.
\end{abstract}
\maketitle

\emph{Introduction.---}Recent experiments try to elucidate the out-of-equilibrium physics of one-dimensional (1D) electron systems \cite{review10}, with experimental systems including quantum wires \cite{barak10}, carbon nanotubes \cite{chen09}, as well as quantum Hall edge channels \cite{granger09,altimiras10}. At low energies, the electron kinetics is dominated by processes within the electron liquid, and the kinetics in 1D is quite distinct from that in higher dimensions \cite{lunde07,gutman09,bagrets09,micklitz10, karzig10, lunde10, micklitz11}. The differences appear already in the most elementary process, namely the accommodation of an additional electron with well-defined energy and momentum which is injected into the liquid. In higher dimensions, the energy and momentum are transferred to a quasiparticle of the Fermi liquid, while the injected charge spreads away from the injection point isotropically in space and on a short time scale governed by the (collective) plasmon excitations. In 1D, such a momentum-conserving tunneling process creates an excited state of the liquid, involving correlated multiple electron-hole excitations. The description of such a state is quite complex \cite{gutman09} even within the Tomonaga-Luttinger model. That raises the question of finding measurable characteristics which quantify the state of the liquid perturbed by electron injection.

Perhaps the simplest characteristic is the partition ratio $Q_-/Q_+$ of the injected charge $e$. The latter creates two pulses which carry unequal charges, $Q_+$ and $Q_-$, propagating, respectively, in and against the direction of motion of the injected charge \cite{pham00,steinberg08}. In the absence of interactions, the entire injected charge moves in the direction of motion of the injected electron, i.e., $Q_-=0$. In the interacting (Luttinger) liquid, $Q_-/Q_+$ is simply related to the ratio of compressibilities of the liquid with and without interactions, and can be readily obtained from the conservation laws of particle number and momentum which yields $Q_\pm = (1 \pm \kappa)/2$ in units of $e$ \cite{review10}. [Here, the Luttinger liquid parameter $\kappa$ measures the interaction strength, with $\kappa =1$ ($\kappa<1$) for non-interacting (repulsively interacting) particles.] The two pulses propagate freely unless they encounter an inhomogeneity of the interaction constant \cite{safi95,maslov95}. Unfortunately, such inhomogeneities are inevitable in experiment which probe the Luttinger liquid by attaching Fermi liquid leads. Because of multiple scattering at the two interfaces, the net charges $ Q_L$ and $Q_R$ flowing into left and right leads differ from the intrinsic values $Q_-$ and $Q_+$. Indeed, $Q_L=0$ in the case of Fermi liquid leads, rendering interaction effects in the Luttinger liquid irrelevant for the charge partitioning measured in $dc$ experiments \cite{steinberg08,berg09}.

The energy of the injected electron is another conserved quantity in the tunneling process which plays a crucial role in the non-equilibrium physics of the electron liquid. In this paper, we show that the energy is also partitioned between left- and right-moving excitations, in a way which is quite distinct from the partitioning of the injected charge and which sensitively probes the interaction strength. When momentum is conserved in the injection process, the initial splitting of the excess energy (measured from the Fermi energy) depends on both energy and momentum of the injected electron as well as the interaction strength $\kappa$. The actual amounts of energy deposited into the two Fermi-liquid leads depend in general on the nature of the interface between Luttinger liquid and leads. The interface is transparent to the flow of energy at high energies, and has finite transparency in the opposite limit. In both limits, the partition of energy deposited in the two leads becomes independent of the properties of the interface but remains a function of $\kappa$ and excess energy. We suggest relatively simple $dc$ experiments to detect energy partitioning and also extend our considerations to include energy partitioning in tunneling into quantum Hall edge states. 
 
\emph{Energy currents in Luttinger liquids.---}We consider a Luttinger liquid of spinless fermions at zero temperature. Decomposing the Luttinger-liquid displacement and phase fields $\phi$ and $\theta$ into right- and left-moving excitations $\theta_\pm(x)=\theta(x) \pm \phi(x)/\kappa$, the relevant Hamiltonian takes the form \cite{pham00}
\begin{equation}
 H=\frac{v_F}{4\pi}\int {\rm d}x \sum_{\alpha=\pm}(\nabla \theta_\alpha)^2 
\end{equation}
with commutation relations $[\theta_\alpha(x),\theta_{\alpha^\prime}(x^\prime)] = \delta_{\alpha\alpha^\prime}  (i\pi\alpha/\kappa)\, {\rm sgn}(x-x^\prime)$. 

\begin{figure}
 \includegraphics[scale=0.5]{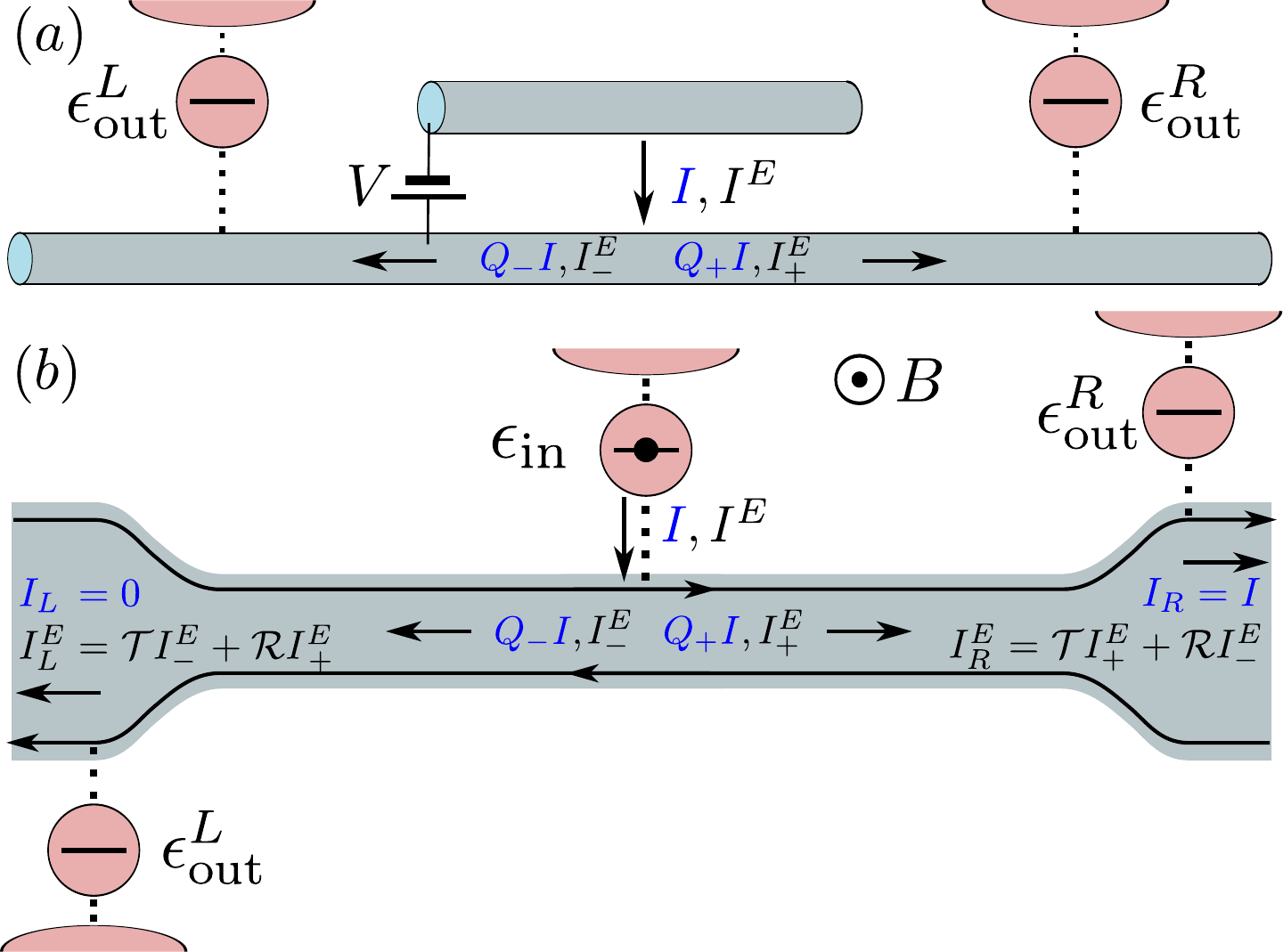}
 \caption{Illustration of proposed experimental setups. (a) Nonlocal injection by momentum-conserving tunneling between parallel quantum wires. The quantum dots to the left and right of the injection region serve to probe the energy partitioning. (b) Local injection into one of two closeby quantum Hall edge channels. The figure indicates both the initial splitting of charge and energy at injection and the resulting splitting in the Fermi-liquid leads. While the charge partitioning is identical for both setups, the energy partitioning is different and distinct from the charge partitioning.}
 \label{fig1}
\end{figure}

\begin{figure}
 \includegraphics[scale=0.4]{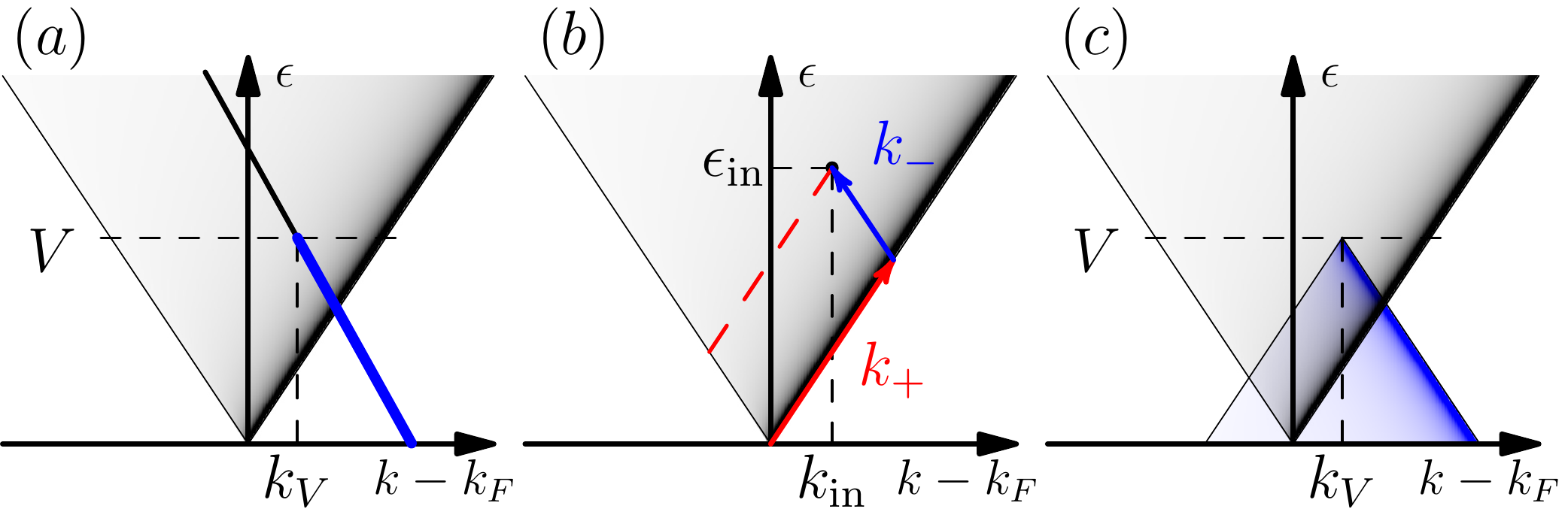}
 \caption{Illustration of nonlocal injection process. (a) Overlap of occupied states in the (non-interacting) source wire (thick blue line) and the Luttinger liquid, as described by the spectral function. The difference between the Fermi energies of source and Luttinger liquid is controlled by the voltage $V$. The Luttinger-liquid spectral function is indicated as a gray-scale background. (b) Illustration of the energy-conservation argument for energy partitioning. (c) For an interacting source, the tunneling current is determined by the overlap of the spectral functions of source and wire.}
 \label{fig2}
\end{figure}

We first consider tunneling from a parallel source wire of length $L_S$ [nonlocal injection, cp.\ Fig.\ \ref{fig1}(a)]. In this case, the dispersions of quantum wire and source can be shifted relative to each other in momentum by applying a magnetic field and in energy by applying a bias voltage $V$ \cite{barak10}. Following recent experiments \cite{barak10}, we assume that these shifts are such that tunneling is only allowed for left movers from the source [field operator $\psi_S(x)$] which tunnel into right-moving free-electron states in the quantum wire [field operator $\psi_R^\dagger(x)$]. This is described by the tunneling Hamiltonian $H_{TR}=t\int_{S}{\rm d}x [\psi_{R}^{\dagger}(x)\psi_{S}(x)+\psi_{S}^{\dagger}(x)\psi_{R}(x)]$ where the nature of the chirality of the states is included through the dispersions, cf.\ Fig.\ \ref{fig2}(a). 

The ensuing right- and left-moving energy currents $I^E_\pm$ in the Luttinger liquid are now described by the operators 
\begin{eqnarray}
I_\pm^{E}\! & = & \!\mathrm{i}[H_{TR},\frac{v_{F}}{4\pi}\int\mathrm{d}x\left(\nabla\theta_\pm\right)^{2}]  
\nonumber \\
&\! = & \!\frac{\pm c t}{2 {\rm i}}Q_\pm \! \int_S \! {\rm d} x  (\!\{\psi_R^\dagger(x),\nabla \theta_\pm(x) \}\psi_S(x)-\rm{h.c.}\! ) 
\label{energycurrentop} 
\end{eqnarray}
To leading order in the tunneling, the expectation value of $I_\pm^E$ becomes
\begin{equation}
 \left\langle I_{\pm}^{E}(\tau)\right\rangle=-\mathrm{i}\int_{-\infty}^{\tau}\mathrm{d}t' \left\langle\left[ I_{\pm}^{E}(\tau),H_{TR}(t')\right]\right\rangle\,.
\label{avIE}
\end{equation}
The resulting correlators can be efficiently computed by writing $\psi_{R}^{\dagger}\sim\mathrm{e}^{-\mathrm{i} \left(Q_{+}A_{+}\nabla \theta_{+}+Q_{-}A_{-}\nabla\theta_{-}\right)}$, expressing them in terms of formal derivatives with respect to the auxiliary operators $A_{\pm}=\nabla^{-1}$, and tracing the modifications due to $A_\pm$ in the standard calculation  \cite{giamarchi} of the Luttinger liquid Green function. 
We then find  
\begin{widetext}
\begin{eqnarray}
\left\langle I_{\pm}^{E}\right\rangle   =  \frac{Q_{\pm}^{2}t^{2} L_{S}}{\kappa}\int \frac{{\mathrm d}\epsilon_S} {2\pi}\int \frac{\mathrm{d}k}{2\pi} \int_{0}^{\infty}{\mathrm{d} \omega_q} \Big\{ G^>_{R,k\mp q}(\epsilon_S  - \omega_{q}) G_{S,k}^<(\epsilon_S-eV)   
 +G^<_{R,k \pm q}(\epsilon_S  + \omega_q ) G^>_{S,k}(\epsilon_S-eV)  \Big\}\, .
\label{parallelwire}
\end{eqnarray}
\end {widetext}
Here, $G^{<,>}_{R,k}(\epsilon)$ denotes the lesser ($<$) or larger ($>$) Green function of the right-moving electrons (with chemical potential $\mu=0$), $G^{<,>}_{S,k}(\epsilon)$ the corresponding Green functions of the left-movers in the source (with chemical potential $\mu_S=eV$), and $\omega_q=cq$ is the plasmon dispersion. The two terms in Eq.\ (\ref{parallelwire}) describe spontaneous plasmon emission in the course of tunneling from source to wire and vice versa, yielding a zero-temperature energy current which is strictly positive. 

A complementary experimental setup would consist of two quantum Hall edge channels spaced such that there is appreciable Coulomb interaction but negligible interedge tunneling. This system shares the same interaction physics with the quantum wire \cite{berg09}, but allows for locally injecting electrons of fixed chirality and fixed energy $\epsilon_{\rm in}$ by selective tunneling into one of the edge channels from a nearby single-level quantum dot [local injection, cp.\ Fig.\ \ref{fig1}(b)]. For tunneling into right-moving states, the tunneling Hamiltonian takes the form $H_{TR} = t_{\rm loc}[\psi_R^\dagger(x=0)\psi_S + {\rm h.c.}]$. Focusing on tunneling from the quantum dot into the quantum wire, i.e., on voltages for which the quantum dot is occupied and described by the Green function $G_S^<(k,\epsilon)=2\pi i \delta(\epsilon + eV - \epsilon_{\rm in})$, we obtain 
\begin{equation}
 \left\langle I_{\pm}^{E}\right\rangle=\frac{i Q_{\pm}^{2} t_{\rm loc}^{2}}{\kappa}\int_{0}^{\infty}\mathrm{d}\omega 
 \int \frac{{\rm d}k}{2\pi} G^>_{R,k}(\epsilon_{\rm in}-\omega).
 \label{IElocal}
\end{equation}
for the left- and right-moving energy currents. 

It is instructive to compare these results for the energy current to the charge current. Charge partitioning is already evident from the {\it operator} relation $I_\pm=Q_{\pm} I$ between the right- and left-moving charge currents $I_\pm=\frac{d}{dt}\{\pm(\kappa/2\pi) \int dx \nabla \theta_\pm\}$ and the total current operator $I=I_++I_-$ \cite{pham00}. Thus, charge partitioning depends only on the interaction parameter and is independent of the particular tunneling process. In contrast, energy partitioning generally depends on the energy and momentum of the tunneling electron which requires one to go beyond the operator level and which makes it sensitive to details of the tunneling process, as we will now discuss in detail. 

\emph{Energy partitioning.---}Focus first on the case of nonlocal injection [Fig.\ \ref{fig1}(a)]. The energetics of the tunneling process from a noninteracting source wire is illustrated in Fig.\ \ref{fig2}(a). While electrons with a distribution of energies and momenta can tunnel into the lower wire, it is easy to separate out the contribution of electrons of well-defined energy and momentum by measuring the differential energy currents ${\rm d}\langle I^E_\pm\rangle/{\rm d}V$. Indeed, Eq.\ (\ref{parallelwire}) yields
\begin{equation}
 \frac{{\rm d} \langle I_{\pm}^E \rangle/{\rm d} V}{{\rm d} \langle I \rangle/{\rm d} V}=\frac{1}{2} (eV \pm c k_V)\,,
\label{diffInonlocal}
\end{equation}
where $V$ is the applied bias and $k_V$ the momentum of the highest-energy electron in the source, cp.\ Fig.\ \ref{fig2}(a). For the double-wire geometry \cite{barak10}, the change in $V$ must be accompanied by an adjustment in the magnetic field such that the crossing of source dispersion and Luttinger liquid mass shell (injection point when both source and wire are noninteracting) remains fixed.

To obtain Eq.\ (\ref{diffInonlocal}), we assume the bias to be such that electrons are tunneling from the source into the lower wire, i.e., we can restrict attention to the first term in Eq.\ (\ref{parallelwire}). Then, Eq.\ (\ref{diffInonlocal}) follows by using the Luttinger-liquid spectral function \cite{giamarchi} (for right- and left-movers) $A_{R/L} = (2\pi/\phi\Gamma^2(\phi))(\Lambda/2c)^{2\phi}|\omega\mp ck|^{\phi-1} |\omega \pm ck|^\phi \theta(|\omega| - c |k|)$ as well as the relation $G^>_{k}(\epsilon) = - i A(k,\epsilon)[1- n_F(\epsilon)]$. Here, $\phi = (\kappa -2 + \kappa^{-1})/4$, $\Lambda$ denotes a large-momentum cutoff, and $n_{F}(\epsilon)$ is the Fermi function.

In essence, this result for energy partitioning can be understood from energy and momentum conservation. A right-moving electron with wavevector $k_F=m v_F$ injected into the Luttinger liquid causes right- and left-moving charge excitations of charges $Q_\pm$ moving with velocity $\pm c$. Since charge transport is accompanied by mass transport, momentum and charge conservation imply $Q_+ m c-Q_- mc=mv_F$ and $Q_+ + Q_- = 1$. This immediately fixes \cite{review10} the charge partitioning $Q_\pm$. Now, consider injection of an electron above the Fermi energy, with energy $\epsilon_F+\epsilon_{\rm in}$ and momentum $k_F+k_{\rm in}$. While the argument for the charge currents remains untouched, conservation of the excess energies and momenta requires
\begin{eqnarray}
 \epsilon_{\rm in}  =  c |k_+| + c |k_-| \,\,\, ; \,\,\,  k_{\rm in}  =  k_+ + k_- \label{con2}\,.
\end{eqnarray}
Here, $k_\pm$ denotes the excess momenta of the left- and right-moving excitations [see Fig.\ \ref{fig2}(b)]. In this way, we find the corresponding excess energies\begin{equation}
\epsilon_\pm=c |k_\pm| = (\epsilon_{\rm in} \pm c k_{\rm in})/2\,, 
\label{energycon}
\end{equation}
which explains Eq.\ (\ref{diffInonlocal}). This result implies that the energy partitioning is entirely independent of the charge partitioning and can be tuned to arbitrary values by varying experimental parameters. In fact, when the momentum of the injected right-moving electron is smaller than the Fermi momentum [$k_{\rm in}<0$, cf.\ Fig \ref{fig2}(b)] , and its energy close to $c|k|$, Eq.\ (\ref{energycon}) implies that essentially all its excess energy is propagating to the left, while most of the charge moves to the right. A crucial ingredient in this result is the interaction-induced broadening of the Luttinger-liquid spectral function which allows for injection of particles away from the mass shell.

For local injection, Eq.\ (\ref{IElocal}) implies 
\begin{equation}
\left\langle I_{\pm}^{E}\right\rangle = \frac{Q_\pm^2} {Q_+^2+Q_-^2} \left\langle I^E\right\rangle
 \label{local}
\end{equation} 
in terms of the total energy current $\left\langle I^E\right\rangle = \epsilon_{\rm in} \langle I \rangle$. Unlike for nonlocal injection, this energy partitioning depends only on the interaction constant, but it is still distinctly different from the charge splitting. This difference can be traced to the fact that the charge density is linear in the Luttinger liquid fields, while the energy is quadratic. 

\emph{Experimental consequences.---}We now turn to experimental signatures of energy partitioning, emphasizing that unlike charge partitioning, it is not masked by the presence of Fermi-liquid leads. For nonlocal injection, the right- and left-moving charge excitations have different maximal energies, given by $\epsilon_{\rm max}^R=(1/2)(eV+ck_V)$ and $\epsilon_{\rm max}^L=(1/2)(eV-ck_V)$ when injecting right-movers. Here, we assume for definiteness that the source wire has a larger charge velocity. Note that these maximal energies remain valid even for an interacting source, cf.\ Fig.\ \ref{fig2}(c). These results can be tested experimentally in some detail in the setup sketched in Fig.\ \ref{fig1}(a), in which the Luttinger liquid is probed by single-level quantum dots both to the left and to the right of the injection region (cp.\ \cite{takei10}). First consider a long Luttinger liquid in the absence of Fermi-liquid leads. In this case, the maximal energies of right- and left-moving excitations are directly observable as thresholds in the current flowing into the quantum dots. Indeed, current can flow into the quantum dots with gate-tunable dot level $\epsilon_{\rm out}^{R/L}$ only as long as $\epsilon_{\rm out}^{R/L} < \epsilon_{\rm max}^{R/L}$. 

In the vicinity of the threshold, the charge currents into the quantum dots will exhibit a power-law dependence on $\epsilon_{\rm max}^{R/L} - \epsilon_{\rm out}^{R/L}$. Extending the approach of Ref.\ \cite{takei10} to the nonlocal injection of electrons of definite chirality, we find for the injection of right movers that
\begin{eqnarray}
dI_R/dV &\propto& (\epsilon_{\rm max}^{R} - \epsilon_{\rm out}^{R})^{\phi-1}
\\
dI_L/dV &\propto& (\epsilon_{\rm max}^{L} - \epsilon_{\rm out}^{L})^{3\phi-\sqrt{\phi(\phi-1)}},
\end{eqnarray}
where $I_{R/L}$ denote the charge currents into right and left quantum dot. These results are valid for a noninteracting source. The expressions for the current in the case of an interacting source are more involved \cite{unpublished}.

In the presence of Fermi-liquid leads, their interface with the Luttinger liquid causes reflection of the energy currents which depends sensitively on the energy $\epsilon$ of the excitations. One may model the interface by $\kappa$ which varies spatially (over a length $d$) from its nominal value in the Luttinger liquid to $\kappa=1$ in the lead. For low energies, $\epsilon \ll  c/d$, the interface can be viewed as abrupt, and the reflection of the energy current is, in close analogy with the Fresnel equations of optics, given by $R_E=1-T_E=(c-v_F)^2/(c+v_F)^2$ \cite{gutman09,footnote}. For larger energies, $\epsilon \gg  c/d$, the interface becomes smooth and reflection of the energy current is exponentially suppressed. Thus, when $\epsilon_{\rm max}^{R/L} \ll  c/d$, there will be multiple reflection of energy currents. In this case, only the larger of the two thresholds can be directly probed experimentally. However, when the threshold energies are sufficiently large, $\epsilon_{\rm max}^{R/L} \gg  c/d$, energy reflection at the interfaces becomes negligible and both thresholds are directly accessible. 

While we considered the spin-polarized case above, the presence of thresholds carries over to the case of a spin-degenerate system supporting spinon excitations. For a linear spectrum with SU(2) symmetry, the injected right mover can excite both left and right-moving charge (with velocity $c_\rho$) modes but only right-moving spin modes (with velocity $c_s$). Consider first the region with $eV > c_\rho |k_V|$. Then, the right threshold $\epsilon^R_{\rm max}$ remains the same as in the spinless case (with $c \to c_\rho$) while the left threshold becomes $\epsilon_{\rm max}^{L} = c_\rho (eV - c_s k_V)/(c_\rho+c_s)$. At lower voltages, $c_\rho |k_V| > eV > c_s |k_V|$, there is no tunneling for $k_V < 0$, while we find $\epsilon_{\rm max}^R = \max\{c_s(eV+c_\rho k_V)/(c_\rho+c_s) ; c_\rho(eV-c_s k_V)/(c_\rho-c_s)\}$ and $\epsilon_{\rm max}^L = c_\rho (eV - c_s k_V)/(c_\rho+c_s)$ for $k_V>0$. 

For local injection into a quantum-Hall edge channel, the thresholds for electron extraction are equal on both sides of the injection point, but the overall right- and left-moving energy currents are different. This remains true after multiple reflections from the Luttinger-liquid-lead interfaces although these reflections affect the overall energy current flowing into the left and right leads. Assuming that the injection energies are sufficiently small such that the Luttinger-liquid-lead interfaces can be treated as abrupt, the energy currents flowing into the right and left leads would be ${\cal T} I_+^E + {\cal R} I_-^E$ and ${\cal R} I_+^E + {\cal T} I_-^E$, respectively. Here, we define ${\cal T} = 1/(1+R_E)$ and ${\cal R} = R_E/(1+R_E)$. These energy currents can in principle be measured directly by probing the electron distribution functions in the outgoing edge channels of the leads (cp., Refs.\ \cite{chen09}). 

\emph{Conclusions.---}While energy and charge of an injected electron travel together in a non-interacting system, this is no longer the case in the presence of interactions. The decoupling caused by interactions is peculiar in one dimension, where it is impossible to separate the excitations into  plasmons and Fermi-liquid quasiparticles. In the Luttinger liquid picture, interactions leave the {\it dc} conductance unchanged \cite{safi95,maslov95}, while significantly affecting, e.g., the thermal conductance \cite{kane96,fazio98,gutman09,footnote}. The decoupling leads to particularly striking consequences when injecting electrons with fixed chirality into a 1d electron system where one may reach conditions such that charge and energy of an injected particle propagate in directions opposite to each other. Finally, energy partitioning is accessible experimentally
with existing abilities and unlike charge partitioning, is detectable in $dc$ setups which include Fermi-liquid leads. 

We acknowledge discussions with G.\ Barak, A.\ Levchenko, T.\ Micklitz, and A.\ Yacoby, financial support through DFG SPP 1538 (FUB), DOE Contract No. DE-FG02-08ER46482 (Yale), and the Packard Foundation (Caltech), as well as the hospitality of the Aspen Center for Physics during part of this work.


\begin{thebibliography}{12}
\bibitem{footnote} This implies that the thermal conductance $G_T$ of a Luttinger liquid depends on the interaction strength $\kappa$. Multiple reflections from the lead-Luttinger-liquid interface give rise to an overall energy transmission of $T_E^{\rm ( tot)} = 2\kappa/(1+\kappa^2)$ and thus $G_T=(\pi T/6)T_E^{\rm (tot)}$, cp.\ Ref.\ \cite{gutman09}.
\end{thebibliography}
\end{document}